\documentclass{revtex4}
\usepackage{amsmath}

\input{tcilatex}

\begin{document}

\title[ ]{d-dimensional non-asymptotically flat thin-shell wormholes in
Einstein-Yang-Mills-Dilaton gravity}
\author{S. Habib Mazharimousavi}
\email{habib.mazhari@emu.edu.tr}
\author{M. Halilsoy}
\email{mustafa.halilsoy@emu.edu.tr}
\author{Z. Amirabi}
\email{zahra.amirabi@emu.edu.tr}
\affiliation{Department of Physics, Eastern Mediterranean University, G. Magusa, north
Cyprus, Mersin 10, Turkey. Tel: 0090 392 6301067, Fax: 0090 3692 365 1604.}

\begin{abstract}
Thin-shell wormholes in Einstein-Yang-Mills-dilaton (EYMD) gravity are
considered. We show that a non-asymptotically flat (NAF) black hole solution
of the d-dimensional EYMD theory provides stable thin-shell wormholes which
are supported entirely by exotic matter. The presence of dilaton makes the
spacetime naturally NAF, and with our conclusion it remains still open to
construct wormholes supported by normal matter between two such spacetimes.
\end{abstract}

\maketitle

\section{INTRODUCTION}

The original aim of a spacetime wormhole was to connect two distinct,
asymptotically flat (AF) spacetimes, or two distant regions in the same AF
spacetime \cite{1,2}. In making such a short cut travel possible it is
crucial that the traveller doesn't encounter event horizons of black holes.
A thin-shell may support such a wormhole provided it has the proper source
to resist against the gravitational collapse. An exotic matter, which fails
to satisfy the energy conditions has been used extensively to provide
maintenance of such wormholes. The non-physical source of energy is confined
on rather thin spherical shells, simply to invoke justification from quantum
theory. More recently, however, it has been shown that without such resort,
thin-shell wormholes can be constructed entirely from normal matter obeying
the energy conditions \cite{3}. Further, such wormholes established on
realistic matter may be stable against radial, linear perturbations. Clearly
this implies that existence of wormholes may be an undeniable reality in our
universe.

In $4-$dimensional (4d) Einstein-Maxwell (EM) theory it was not possible to
employ normal matter in the construction / maintenance of thin-shell
wormholes. In 5d, with the Gauss-Bonnet (GB) extension of EM theory \cite{4}%
, it was further proved that stable, thin-shell wormholes supported by
normal matter is possible provided the GB parameter takes negative values,
i.e. $\alpha <0$ \cite{3}$.$ Is this true also for different sources such as
Yang-Mills (YM) fields when considered in Einstein-Gauss-Bonnet (EGB)
gravity? The answer, to the best of our knowledge, is not in the affirmative.

In this Letter we employ dilaton field beside YM field to investigate the
reality of such thin-shell wormholes. In doing this it should be remembered
that the strong coupling of dilaton turns the spacetime into a
non-asymptotically flat (NAF) one. This is a digression from the original
idea of a wormhole since we have to revise the advantages of an AF
spacetime. We remind, however, that we have already enough familiarity with
the cases of NAF spacetimes, the best known one being the de (anti)-Sitter.
For this reason we extend the concept of a wormhole in an AF spacetime to a
NAF one through the prominent source of a dilaton. Such a wormhole will
provide traversability from one NAF to another NAF spacetime. For this
purpose we employ d-dimensional solutions of Einstein-Yang-Mills-dilaton
(EYMD) theory found recently \cite{5,6}, and investigate thin-shell
construction in such geometries. It turns out that in the dilatonic
thin-shells the required source must be exotic. Our findings show, against
our expectation that neither the YM charge, nor the dilatonic parameter have
significant effect on the negative energy density of the thin-shell. A
deeper reasoning should be sought in the Kaluza-Klein reduction procedure in
which dilaton is created from higher dimensional EM theory. The latter needs
also an exotic matter to support and maintain a thin-shell wormhole. To have
anything but exotic, with reference to our past experience, the dilaton must
vanish and additional geometrical structures, such as Gauss-Bonnet (GB)
terms must be added.

The Letter is organized as follows. In Sec. II we give a brief review of the
d-dimensional EYMD gravity. Dynamic thin-shell wormhole construction and
stability analysis is presented in Sec. III. The paper ends with our
conclusion, which appears in Sec. IV.

\section{A BRIEF REVIEW OF HIGHER DIMENSIONAL\ EYMD GRAVITY}

We consider the $d-$dimensional action in the EYMD theory as $(G=1)$%
\begin{eqnarray}
S &=&-\frac{1}{16\pi }\int\nolimits_{\mathcal{M}}d^{d}x\sqrt{-g}\left( R-%
\frac{4}{d-2}\left( \mathbf{\nabla }\Phi \right) ^{2}+\mathcal{L}\left( \Phi
\right) \right) ,\text{ \ \ }  \notag \\
\mathcal{L}\left( \Phi \right) &=&-e^{-4\alpha \Phi /\left( d-2\right) }%
\mathbf{Tr}(F_{\lambda \sigma }^{\left( a\right) }F^{\left( a\right) \lambda
\sigma }),
\end{eqnarray}%
where%
\begin{equation}
\mathbf{Tr}(.)=\tsum_{a=1}^{6}\left( .\right) ,
\end{equation}%
$\Phi $ is the dilaton scalar potential, the parameter $\alpha $ denotes the
coupling between dilaton and Yang-Mills (YM) field and as usual $R$ is the
Ricci scalar. The YM field $2-$forms $\mathbf{F}^{\left( a\right) }=F_{\mu
\nu }^{\left( a\right) }dx^{\mu }\wedge dx^{\nu }$ are given by \cite{5,6}%
\begin{equation}
\mathbf{F}^{\left( a\right) }=\mathbf{dA}^{\left( a\right) }+\frac{1}{%
2\sigma }C_{\left( b\right) \left( c\right) }^{\left( a\right) }\mathbf{A}%
^{\left( b\right) }\wedge \mathbf{A}^{\left( c\right) }
\end{equation}%
in which $C_{\left( b\right) \left( c\right) }^{\left( a\right) }$ are the
structure constants, $\sigma $ is a coupling constant and the YM potential $%
1-$forms are given by%
\begin{equation}
\mathbf{A}^{(a)}=\frac{Q}{r^{2}}\left( x_{i}dx_{j}-x_{j}dx_{i}\right) ,\text{
\ \ }Q=\text{charge, \ }r^{2}=\overset{d-1}{\underset{i=1}{\sum }}x_{i}^{2},%
\text{ \ }2\leq j+1\leq i\leq d-1,\text{ \ and \ }1\leq a\leq \left(
d-1\right) (d-2)/2,
\end{equation}%
which is the higher dimensional Wu-Yang ansatz \cite{5,6}. The field
equations, after varying the action, are given by%
\begin{equation}
\mathbf{d}\left( e^{-4\alpha \Phi /\left( d-2\right) \star }\mathbf{F}%
^{\left( a\right) }\right) +\frac{1}{\sigma }C_{\left( b\right) \left(
c\right) }^{\left( a\right) }e^{-4\alpha \Phi /\left( d-2\right) }\mathbf{A}%
^{\left( b\right) }\wedge ^{\star }\mathbf{F}^{\left( c\right) }=0,
\end{equation}%
\begin{equation}
R_{\mu \nu }=\frac{4}{d-2}\mathbf{\partial }_{\mu }\Phi \partial _{\nu }\Phi
+2e^{-4\alpha \Phi /\left( d-2\right) }\left[ \mathbf{Tr}\left( F_{\mu
\lambda }^{\left( a\right) }F_{\nu }^{\left( a\right) \ \lambda }\right) -%
\frac{1}{2\left( d-2\right) }\mathbf{Tr}(F_{\lambda \sigma }^{\left(
a\right) }F^{\left( a\right) \lambda \sigma })g_{\mu \nu }\right] ,
\end{equation}%
\begin{equation}
\nabla ^{2}\Phi =-\frac{1}{2}\alpha e^{-4\alpha \Phi /\left( d-2\right) }%
\mathbf{Tr}(F_{\lambda \sigma }^{\left( a\right) }F^{\left( a\right) \lambda
\sigma }),
\end{equation}%
in which $R_{\mu \nu }$ is the Ricci tensor and the hodge star $^{\star }$
means duality. As it was shown in Ref. \cite{5,6}, these equations admit
black hole solution in the form of 
\begin{equation}
ds^{2}=-f\left( r\right) dt^{2}+\frac{dr^{2}}{f\left( r\right) }+h\left(
r\right) ^{2}d\Omega _{d-2}^{2},
\end{equation}%
where $d\Omega _{d-2}^{2}$ is the line element on $S^{d-2}$ and the solution
can be summarized as follows 
\begin{equation}
\Phi =-\frac{\left( d-2\right) }{2}\frac{\alpha \ln r}{\alpha ^{2}+1},\text{
\ \ }h\left( r\right) =Ar^{\frac{\alpha ^{2}}{\alpha ^{2}+1}},\text{ \ \ }%
f\left( r\right) =\Xi \left( 1-\left( \frac{r_{h}}{r}\right) ^{\frac{\left(
d-3\right) \alpha ^{2}+1}{\alpha ^{2}+1}}\right) r^{\frac{2}{\alpha ^{2}+1}}.
\end{equation}%
We abbreviate here%
\begin{equation}
\Xi =\frac{\left( d-3\right) }{\left( \left( d-3\right) \alpha ^{2}+1\right)
Q^{2}},\text{ \ \ }A^{2}=Q^{2}\left( \alpha ^{2}+1\right) ,\text{ }%
r_{h}=\left( \frac{4\left( \alpha ^{2}+1\right) M}{\left( d-2\right) \Xi
\alpha ^{2}A^{d-2}}\right)
\end{equation}%
and $r_{h}$ stands for the radius of event horizon. Here $M$ implies the
quasilocal mass (see \cite{6} and the references therein).

\section{DYNAMIC THIN-SHELL WORMHOLES IN EYMD GRAVITY}

Consider two copies of the EYMD spacetime 
\begin{equation}
M^{\pm }=\left\{ r^{_{\pm }}\geq a,\text{ \ }a>r_{h}\right\}
\end{equation}%
and paste them at the boundary hypersurface $\Sigma ^{\pm }=\left\{ r^{_{\pm
}}=a,\text{\ }a>r_{h}\right\} $. These surfaces are identified on $r=a$ with
a surface energy-momentum of a thin-shell such that geodesic completeness
holds. Following the Darmois-Israel formalism \cite{7} in terms of the
original coordinates $x^{\gamma }=\left( t,r,\theta _{1},\theta
_{2},...\right) $ (i.e. on $M$) the induced metric $\xi ^{i}=\left( \tau
,\theta _{1},\theta _{2},...\right) $ on $\Sigma $ is given by (Latin
indices run over the induced coordinates i.e., $\left\{ 1,2,...,d-1\right\} $
and Greek indices run over the original manifold's coordinates i.e., $%
\left\{ 1,2,...,d\right\} $)%
\begin{equation}
g_{ij}=\frac{\partial x^{\alpha }}{\partial \xi ^{i}}\frac{\partial x^{\beta
}}{\partial \xi ^{j}}g_{\alpha \beta }.
\end{equation}%
We shall study now the dynamic thin-shell wormhole by letting the throat
radius to be a function of proper time $\tau $, i.e. $a=a\left( \tau \right) 
$. ( A $4-$dimensional formalism similar to these section can be found in
Ref.\cite{8}). One can easily show that the induced metric is in the form%
\begin{equation}
g_{ij}=\text{diag}\left( -1,h\left( a\left( \tau \right) \right)
^{2},h\left( a\left( \tau \right) \right) ^{2}\sin ^{2}\theta _{1},h\left(
a\left( \tau \right) \right) ^{2}\sin ^{2}\theta _{1}\sin ^{2}\theta
_{2},....\right) .
\end{equation}%
The parametric equation of the hypersurface $\Sigma $ is given by 
\begin{equation}
F\left( r,a\left( \tau \right) \right) =r-a\left( \tau \right) =0,
\end{equation}%
and the normal unit vectors to $M^{\pm }$ defined by 
\begin{equation}
n_{\gamma }=\left( \pm \left\vert g^{\alpha \beta }\frac{\partial F}{%
\partial x^{\alpha }}\frac{\partial F}{\partial x^{\beta }}\right\vert
^{-1/2}\frac{\partial F}{\partial x^{\gamma }}\right) _{r=a},
\end{equation}%
are found as follows%
\begin{equation}
n_{t}=\pm \left( \left\vert g^{tt}\left( \frac{\partial a\left( \tau \right) 
}{\partial t}\right) ^{2}+g^{rr}\right\vert ^{-1/2}\frac{\partial F}{%
\partial t}\right) _{r=a}.
\end{equation}%
Upon using%
\begin{equation}
\left( \frac{\partial t}{\partial \tau }\right) ^{2}=\frac{1}{f\left(
a\right) }\left( 1+\frac{1}{f\left( a\right) }\dot{a}^{2}\right) ,
\end{equation}%
it implies%
\begin{equation}
n_{t}=\pm \left( -\dot{a}\right) .
\end{equation}%
Similarly one finds that%
\begin{equation}
n_{r}=\pm \left( \left\vert g^{tt}\frac{\partial F}{\partial t}\frac{%
\partial F}{\partial t}+g^{rr}\frac{\partial F}{\partial r}\frac{\partial F}{%
\partial r}\right\vert ^{-1/2}\frac{\partial F}{\partial r}\right)
_{r=a}=\pm \left( \frac{\sqrt{f\left( a\right) +\dot{a}^{2}}}{f\left(
a\right) }\right) ,\text{ and }n_{\theta _{i}}=0,\text{ for all }\theta _{i}.
\end{equation}%
After the unit $d-$normal, one finds the extrinsic curvature tensor
components from the definition%
\begin{equation}
K_{ij}^{\pm }=-n_{\gamma }^{\pm }\left( \frac{\partial ^{2}x^{\gamma }}{%
\partial \xi ^{i}\partial \xi ^{j}}+\Gamma _{\alpha \beta }^{\gamma }\frac{%
\partial x^{\alpha }}{\partial \xi ^{i}}\frac{\partial x^{\beta }}{\partial
\xi ^{j}}\right) _{r=a}.
\end{equation}%
It follows that 
\begin{gather}
K_{\tau \tau }^{\pm }=-n_{t}^{\pm }\left( \frac{\partial ^{2}t}{\partial
\tau ^{2}}+\Gamma _{\alpha \beta }^{t}\frac{\partial x^{\alpha }}{\partial
\tau }\frac{\partial x^{\beta }}{\partial \tau }\right) _{r=a}-n_{r}^{\pm
}\left( \frac{\partial ^{2}r}{\partial \tau ^{2}}+\Gamma _{\alpha \beta }^{r}%
\frac{\partial x^{\alpha }}{\partial \tau }\frac{\partial x^{\beta }}{%
\partial \tau }\right) _{r=a}=  \notag \\
-n_{t}^{\pm }\left( \frac{\partial ^{2}t}{\partial \tau ^{2}}+2\Gamma
_{tr}^{t}\frac{\partial t}{\partial \tau }\frac{\partial r}{\partial \tau }%
\right) _{r=a}-n_{r}^{\pm }\left( \frac{\partial ^{2}r}{\partial \tau ^{2}}%
+\Gamma _{tt}^{r}\frac{\partial t}{\partial \tau }\frac{\partial t}{\partial
\tau }+\Gamma _{rr}^{r}\frac{\partial r}{\partial \tau }\frac{\partial r}{%
\partial \tau }\right) _{r=a}=\pm \left( -\frac{f^{\prime }+2\ddot{a}}{2%
\sqrt{f+\dot{a}^{2}}}\right) ,
\end{gather}%
Also 
\begin{equation}
K_{\theta _{i}\theta _{i}}^{\pm }=-n_{\gamma }^{\pm }\left( \frac{\partial
^{2}x^{\gamma }}{\partial \theta _{i}^{2}}+\Gamma _{\alpha \beta }^{\gamma }%
\frac{\partial x^{\alpha }}{\partial \theta _{i}}\frac{\partial x^{\beta }}{%
\partial \theta _{i}}\right) _{r=a}=\pm \sqrt{f\left( a\right) +\dot{a}^{2}}%
hh^{\prime }.
\end{equation}%
In sum, we have%
\begin{equation}
\left\langle K_{ij}\right\rangle =\text{diag}\left( -\frac{f^{\prime }+2%
\ddot{a}}{\sqrt{f+\dot{a}^{2}}},2\sqrt{f\left( a\right) +\dot{a}^{2}}%
hh^{\prime },2\sqrt{f\left( a\right) +\dot{a}^{2}}hh^{\prime }\sin
^{2}\theta _{1},2\sqrt{f\left( a\right) +\dot{a}^{2}}hh^{\prime }\sin
^{2}\theta _{1}\sin ^{2}\theta _{2},...\right) ,
\end{equation}%
which implies 
\begin{equation}
\left\langle K_{i}^{j}\right\rangle =\text{diag}\left( \frac{f^{\prime }+2%
\ddot{a}}{\sqrt{f+\dot{a}^{2}}},2\sqrt{f\left( a\right) +\dot{a}^{2}}\frac{%
h^{\prime }}{h},2\sqrt{f\left( a\right) +\dot{a}^{2}}\frac{h^{\prime }}{h},2%
\sqrt{f\left( a\right) +\dot{a}^{2}}\frac{h^{\prime }}{h},...\right)
\end{equation}%
and therefore%
\begin{equation}
K=\text{Trace}\left\langle K_{i}^{j}\right\rangle =\left\langle
K_{i}^{i}\right\rangle =\frac{f^{\prime }+2\ddot{a}}{\sqrt{f+\dot{a}^{2}}}%
+2\left( d-2\right) \sqrt{f\left( a\right) +\dot{a}^{2}}\frac{h^{\prime }}{h}%
.
\end{equation}%
The surface energy-momentum components of the thin-shell are \cite{9,10}%
\begin{equation}
S_{i}^{j}=-\frac{1}{8\pi }\left( \left\langle K_{i}^{j}\right\rangle
-K\delta _{i}^{j}\right)
\end{equation}%
which yield%
\begin{gather}
\sigma =-S_{\tau }^{\tau }=-\frac{\left( d-2\right) }{4\pi }\left( \sqrt{%
f\left( a\right) +\dot{a}^{2}}\frac{h^{\prime }}{h}\right) , \\
S_{\theta _{i}}^{\theta _{i}}=p_{\theta _{i}}=\frac{1}{8\pi }\left( \frac{%
f^{\prime }+2\ddot{a}}{\sqrt{f\left( a\right) +\dot{a}^{2}}}+2\left(
d-3\right) \sqrt{f\left( a\right) +\dot{a}^{2}}\frac{h^{\prime }}{h}\right) .
\end{gather}%
By substitution one can show that the energy conservation takes the form 
\begin{equation}
\frac{d}{d\tau }\left( \sigma \mathcal{A}\right) +p\frac{d}{d\tau }\left( 
\mathcal{A}\right) =-\frac{\left( d-2\right) }{4\pi }\frac{h^{\prime \prime }%
}{h}\dot{a}\mathcal{A}\sqrt{f\left( a\right) +\dot{a}^{2}}\neq 0,
\end{equation}%
in which $\mathcal{A}=\frac{2\pi ^{\frac{d-1}{2}}}{\Gamma \left( \frac{d-1}{2%
}\right) }h\left( a\right) ^{d-2}$ is the area of the thin-shell. In other
words, due to the exchange with the bulk spacetime, the energy on the shell
is not conserved.

Let us note that we adopt the junction conditions of general relativity
which can be justified by the fact that the dilaton field $\Phi $ and its
normal derivative are both continuous across the shell. Similar approach has
been followed by different authors \cite{11} which can be justified easily
by integrating the dilaton equation (7) across the throat radius $a_{0}\pm
\epsilon ,$ in the limit as $\epsilon \rightarrow 0.$ Since the singularity
and horizons all reside deliberately at distances $r<a_{0}$, the
contribution from the dilaton field and its derivative to the thin-shell
source vanishes. We note that this is different in the case of Brans-Dicke
(BD) scalar field, which has structural difference compared with the dilaton
field \cite{12}. To say the least among others, the exponential coupling of
dilaton with the gauge field makes it short ranged whereas BD field is long
ranged. To calculate the amount of exotic matter needed to construct the
traversable wormhole we use the integral%
\begin{equation}
\Omega =\int \sqrt{-g}\left( \rho +p_{r}\right) dx^{d-1}.
\end{equation}%
For a thin-shell wormhole $p_{r}=0$ and $\rho =\sigma \delta \left(
r-a\right) ,$ where $\delta \left( r-a\right) $ is the Dirac delta function.
In static configuration, a simple calculation gives 
\begin{equation}
\Omega =\int\limits_{0}^{2\pi }\int\limits_{0}^{\pi }...\int\limits_{0}^{\pi
}\int\limits_{0}^{\infty }\sqrt{-g}\sigma \delta \left( r-a\right) drd\theta
_{1}d\theta _{2}...d\theta _{d-2}=\frac{2\pi ^{\frac{d-1}{2}}}{\Gamma \left( 
\frac{d-1}{2}\right) }h\left( a\right) ^{d-2}\sigma \left( a\right) .
\end{equation}

We aim now to apply a small radial perturbation around the radius of
equilibrium $a_{0}$ and to investigate the behavior of the throat under this
perturbation. For this perturbation we consider the radial pressure of the
thin-shell to be a linear function of the energy density, i.e., \cite{13}%
\begin{equation}
p=p_{0}+\beta ^{2}\left( \sigma -\sigma _{0}\right) .
\end{equation}%
Here $p_{0}$ and $\sigma _{0}$ are the radial pressure and energy density of
the thin-shell in the static configuration of radius $a_{0}$ which are given
by 
\begin{eqnarray}
\sigma _{0} &=&-\frac{\left( d-2\right) }{4\pi }\left( \sqrt{f\left(
a_{0}\right) }\frac{h^{\prime }\left( a_{0}\right) }{h\left( a_{0}\right) }%
\right) , \\
p_{0} &=&\frac{1}{8\pi }\left( \frac{f^{\prime }\left( a_{0}\right) }{\sqrt{%
f\left( a_{0}\right) }}+2\left( d-3\right) \sqrt{f\left( a_{0}\right) }\frac{%
h^{\prime }\left( a_{0}\right) }{h\left( a_{0}\right) }\right) ,
\end{eqnarray}%
and $\beta ^{2}$ is a constant related to the speed of sound. By
substituting (32) into (31), one finds a first order differential equation
for $\sigma \left( a\right) $ which is given by 
\begin{equation}
\sigma ^{\prime }\left( a\right) +\left( d-2\right) \left( \sigma +p\right) 
\frac{h^{\prime }\left( a\right) }{h\left( a\right) }=\frac{h^{\prime \prime
}\left( a\right) }{h^{\prime }\left( a\right) }\sigma \left( a\right) ,
\end{equation}%
or equivalently%
\begin{equation}
\sigma ^{\prime }\left( a\right) +\sigma \left( a\right) \left[ \left(
d-2\right) \left( 1+\beta ^{2}\right) \frac{h^{\prime }\left( a\right) }{%
h\left( a\right) }-\frac{h^{\prime \prime }\left( a\right) }{h^{\prime
}\left( a\right) }\right] =\left( d-2\right) \left( \sigma _{0}\beta
^{2}-p_{0}\right) \frac{h^{\prime }\left( a\right) }{h\left( a\right) }.
\end{equation}

This equation, for the case of EYMD wormhole introduced before can be
expressed as 
\begin{equation}
r\sigma ^{\prime }\left( a\right) +\xi _{1}\sigma \left( a\right) =\xi _{2}
\end{equation}%
in which%
\begin{eqnarray}
\xi _{1} &=&\frac{1+\left( d-2\right) \alpha ^{2}\left( 1+\beta ^{2}\right) 
}{\alpha ^{2}+1}, \\
\xi _{2} &=&\frac{\left( d-2\right) \alpha ^{2}\left( \sigma _{0}\beta
^{2}-p_{0}\right) }{\alpha ^{2}+1}.
\end{eqnarray}%
This equation admits a solution in the form of%
\begin{equation}
\sigma \left( a\right) =\frac{\xi _{2}}{\xi _{1}}+\left( \sigma _{0}-\frac{%
\xi _{2}}{\xi _{1}}\right) \left( \frac{a_{0}}{a}\right) ^{\xi _{1}},
\end{equation}%
which obviously at $a=a_{0}$ is nothing but $\sigma _{0}.$ In terms of the
metric function $f\left( a\right) ,$ in a dynamic case $\sigma \left(
a\right) $ was given by (35) which after equating with the latter expression
we obtain 
\begin{equation}
\dot{a}^{2}+V\left( a\right) =0.
\end{equation}%
Here the potential function $V\left( a\right) $ is given by 
\begin{equation}
V\left( a\right) =f\left( a\right) -\frac{16\pi ^{2}a^{2}}{\left( d-2\right)
^{2}}\left( \frac{\alpha ^{2}+1}{\alpha ^{2}}\right) ^{2}\left[ \frac{\xi
_{2}}{\xi _{1}}+\left( \sigma _{0}-\frac{\xi _{2}}{\xi _{1}}\right) \left( 
\frac{a_{0}}{a}\right) ^{\xi _{1}}\right] ^{2}.
\end{equation}%
We notice that $V\left( a\right) ,$ and more tediously $V^{\prime }\left(
a\right) ,$ both vanish at $a=a_{0}.$ The stability requirement for
equilibrium reduces therefore to the determination of the regions in which$\
V^{\prime \prime }(a_{0})>0.$ For this reason we shall proceed through
numerical analysis to see whether stability regions/ islands develop and
under what conditions. Let us note that our figures refer to $d=5$, for $d>5$
we observed that no significant changes take place. In Fig.s 1-2 we show the
stability regions in terms of $\beta $ and wormhole throat $a_{0}.$ (Note
that we use scalings $\Omega =Q^{2}\tilde{\Omega}$, $f=\frac{\tilde{f}}{Q^{2}%
}$ and $p_{0}=\frac{1}{\left\vert Q\right\vert }\tilde{p}_{0}$ in terms of
the YM charge $Q$ and we plot ($\tilde{\Omega},\tilde{f},\tilde{p}_{0}$)).
We also show $\tilde{\Omega}$ in terms of $a_{0}$ for different values of $%
\alpha $. As stated before, our wormhole is supported entirely by negative
energy, $\tilde{\Omega}<0$ on the shell. For changing dilatonic parameter $%
\alpha $ the change in $\tilde{\Omega}$ is visible in the plots shown. For
fixed mass, decreasing $\alpha $ improves $\tilde{\Omega}$ toward zero line
but yet in the $\left( -\right) $ domain. The pressure on the shell
increases with increasing $\alpha $. It is noticed from the dark stability
regions that no stable region forms for $\left\vert \beta \right\vert <1$,
i.e. below the speed of light. Consideration of dimensions $d>5$, doesn't
change the behaviors of $d=5$ much, this can be seen by further plots which
we shall ignore in this study.

\section{CONCLUSION}

We have introduced $d-$dimensional thin-shell wormholes in EYMD gravity
supported by exotic matter. Although in quantum theory such matter finds
acceptance on reasonable grounds in classical physics we should be more
cautious. By choosing finely-tuned parameters of mass and dilaton parameter $%
\alpha $ we provide the stability of such wormholes against linear, radial
perturbations. For any finite YM charge which amounts to scale the energy
and pressure no significant difference is observed in the plots. It should
be added that extreme YM charge, such as $Q\rightarrow 0(\infty ),$ results
in extreme energy (but still ($-$)) and pressure conditions. When the YM and
dilaton fields vanish we recover the previously known results. Inclusion of
dilaton converts the spacetime from asymptotically flat (AF) to
non-asymptotically flat (NAF) case and only exotic matter can support such a
wormhole. We recall that wormholes in NAF universe (without dilaton) is not
a new idea \cite{14}. Accordingly, the mass is no more that of ADM but
rather the quasi-local mass is implied. Vanishing of dilaton brings us to
EYM theory, which may be supplemented with modified gravity to pave the way
toward wormholes with normal matter. We have shown elsewhere that inclusion
of Gauss-Bonnet (GB) term accomplishes this task and provides stable,
thin-shell wormholes with normal (i.e. non-exotic) matter \cite{15}.

\textbf{Figure\textbf{\ Caption}s:}

Figure 1: The plot of $\ \tilde{f}\left( a_{0}\right) =Q^{2}f\left(
a_{0}\right) ,$ $\tilde{\Omega}\left( a_{0}\right) =\Omega \left(
a_{0}\right) /Q^{2}$ and pressure $\tilde{p}_{0}=p_{0}\left\vert
Q\right\vert $, for $d=5$. The shaded inscribed part shows the stable
regions (i.e. $V^{\prime \prime }\left( a_{0}\right) >0$) in the $\left(
\beta ,a_{0}\right) $ diagram. Fig.1a) for $M=0.1$,$\alpha =2.0,$ Fig.1b)
for $M=0.10$,$\alpha =1.0,$ and Fig. 1c) for $M=0.05$,$\alpha =1.0.$ It is
seen that the $\Omega $ behavior doesn't differ much in this range of
parameters.

Figure 2: Similar plots for $\tilde{f}\left( a_{0}\right) ,$ $\tilde{\Omega}%
\left( a_{0}\right) $ and pressure $\tilde{p}_{0}$, (again for $d=5$), for
2a) $M=0.10$, $\alpha =0.4$ and 2b) $M=0.10$, $\alpha =0.2.$ Decreasing $%
\alpha $ values improves $\tilde{\Omega}\left( a_{0}\right) $ slightly which
still lies in the $(-)$ domain. Smaller $\alpha $ implies also less pressure
($\tilde{p}_{0}$) on the shell. Stable regions ($V^{\prime \prime }\left(
a_{0}\right) >0$) are shown, versus ($\beta ,a_{0}$) in dark.\newpage

\bigskip


\bigskip

\begin{thebibliography}{99}
\bibitem{1} M. S. Morris and K. S. Thorne, Am. J. Phys. \textbf{56}, 395
(1988).

\bibitem{2} M. Visser, Lorantzian Wormholes (AIP Press, Newyork, 1996).

\bibitem{3} M. G. Richarte and C. Simeone Phys. Rev. D \textbf{76}, 087502
(2007); Erratum-ibid \textbf{77}, 089903 (2008)E; S. H. Mazharimousavi, M.
Halilsoy and Z. Amirabi, Phys. Rev. D \textbf{81}, 104002 (2010).

\bibitem{4} B. Zwiebach, Phys. Lett. B \textbf{156}, 315(1985); D. G.
Boulware and S. Deser, Phys. Rev. Lett., \textbf{55}, 2656(1985); R. G. Cai
and K. S. Soh, Phys. Rev. D \textbf{59}, 044013 (1999); R. G. Cai, ibid. 
\textbf{65}, 084014(2002); R. Aros, R. Troncoso and J. Zanelli, Phys. Rev. D 
\textbf{63}, 084015(2001); Y. M. Cho and I. P. Neupane, Phys. Rev. D \textbf{%
66}, 024044(2002); M. H. Dehghani, Phys. Rev. D \textbf{67,} 064017(2003).

\bibitem{5} S. H. Mazharimousavi and M. Halilsoy, Phys. Rev. D \textbf{76},
087501 (2007).

\bibitem{6} S. H. Mazharimousavi, M. Halilsoy and Z. Amirabi, Gen. Relativ.
Gravit. \textbf{42}, 261 (2010).

\bibitem{7} G. Darmois, M\'{e}morial des Sciences Math\'{e}matiques,
Fascicule XXV (Gauthier-Villars, Paris, 1927), Chap. V; W. Israel, Nuovo
Cimento B \textbf{44}, 1(1966); B \textbf{48}, 463(E)(1967); P. Musgrave and
K. Lake, Class. Quant. Grav. \textbf{13}, 1885 (1996).

\bibitem{8} E. F. Eiroa, Phys. Pev. D \textbf{78}, 024018 (2008).

\bibitem{9} N. Sen, Ann. Phys. (Leipzig) 73 (1924) 365; K. Lanczos, Ann.
Phys. (Leipzig) 74 (1924) 518; W. Israel, Nuovo Cimento B 44 (1966) 1; W.
Israel, Nuovo Cimento B 48 (1967) 463, Erratum.

\bibitem{10} S. C. Davis, Phys. Rev. D, \textbf{67}, 024030 (2003); F. S. N.
Lobo, Gen. Rel. Grav. \textbf{37}, 2023 (2005 ); K. K. Nandi, Y. Z. Zhang
and K. B. Vijaya Kumar, Phys. Rev. D \textbf{70}, 127503 (2004).

\bibitem{11} E. F. Eiroa and C. Simeone, Phys. Rev. D \textbf{71}, 127501
(2005);

A. A. Usmani, Z. Hasan, F. Rahaman, Sk. A. Rakib, Saibal Ray and Peter K. F.
Kuhfittig, Gen. Rel. Grav. (2010 ) in press, Doi: 10.1007/s10714-010-1044-y.

\bibitem{12} L. A. Anchordoqui, S. E. Perez Bergliaffa, and D. F. Torres,
Phys. Rev. D \textbf{55}, 5226 (1997);

F. Eiroa, M. G. Richarte, and C. Simeone, Phys. Lett. A \textbf{373} 1
(2008), Erratum-ibid. \textbf{373}, 2399 (2009);

F. Dahia and C. Romero, Phys. Rev. D \textbf{60}, 104019 (1999).

\bibitem{13} P. R. Brady, J. Louko and E. Poisson, Phys. Rev. D \textbf{44},
1891(1991); E. Poisson and M. Visser, Phys. Rev. D \textbf{52}, 7318(1995).

\bibitem{14} O. B. Zaslavskii, Phys. Rev. D 72, 061303(R) (2005).

\bibitem{15} S. Habib Mazharimousavi, M. Halilsoy and Z. Amirabi,
arXiv:1007.4627.
\end{thebibliography}
\end{document}